\documentclass[9pt,sigplan,screen,nonacm]{acmart}

\setcopyright{none}

\usepackage{amsmath,amssymb,amsfonts}
\usepackage{subcaption}
\usepackage{graphicx}
\hypersetup{hidelinks}
\usepackage{xcolor}
\usepackage{enumitem}
\usepackage{comment}
\usepackage{ifthen}
\usepackage{tikz}

\newboolean{showcomments}
\setboolean{showcomments}{true}

\ifthenelse{\boolean{showcomments}}
{ \newcommand{\mynote}[3]{
    \fbox{\bfseries\sffamily\scriptsize#1}
    {\small$\blacktriangleright$\textsf{\emph{\color{#3}{#2}}}$\blacktriangleleft$}}}
{ \newcommand{\mynote}[3]{}}

\definecolor{pink}{rgb}{1,0.2,0.7}
\definecolor{purple}{rgb}{0.7,0,0.9}

\usepackage[most]{tcolorbox}
\newtcolorbox{observationbox}{
  colback=white,
  colframe=black,
  arc=0pt,
  boxrule=0.5pt,
  left=1pt,right=1pt,top=1pt,bottom=1pt
}

\author{I-Ting Lee$^{*}$}
\affiliation{
  \institution{Department of Computer Science and Information Engineering}
  \institution{National Cheng Kung University}
  \city{Tainan}
  \country{Taiwan}
}
\email{0321iting@gmail.com}

\author{Bao-Kai Wang$^{*}$}
\affiliation{
  \institution{Department of Computer Science and Information Engineering}
  \institution{National Cheng Kung University}
  \city{Tainan}
  \country{Taiwan}
}
\email{kk29880015@gmail.com}

\author{Liang-Chi Chen}
\affiliation{
  \institution{Department of Computer Science and Information Engineering}
  \institution{National Taiwan University}
  \city{Taipei}
  \country{Taiwan}
}
\email{d12922012@csie.ntu.edu.tw}

\author{Wen Sheng Lim}
\affiliation{
  \institution{Department of Computer Science and Information Engineering}
  \institution{National Taiwan University}
  \city{Taipei}
  \country{Taiwan}
}
\email{tundergod1882@gmail.com}

\author{Da-Wei Chang}
\affiliation{
  \institution{Department of Computer Science and Information Engineering}
  \institution{National Cheng Kung University}
  \city{Tainan}
  \country{Taiwan}
}
\email{davidchang@csie.ncku.edu.tw}

\author{Yu-Ming Chang}
\affiliation{
  \institution{Wolley}
  \city{Taipei}
  \country{Taiwan}
}
\email{san@wolleytech.com}

\author{Chieng-Chung Ho}
\affiliation{
  \institution{Department of Computer Science and Information Engineering}
  \institution{National Cheng Kung University}
  \city{Tainan}
  \country{Taiwan}
}
\email{ccho@gs.ncku.edu.tw}

\thanks{$^{*}$ Equal contribution.}

\begin{document}

\title{PIM or CXL-PIM? Understanding Architectural Trade-offs Through Large-Scale Benchmarking}

\begin{abstract}
Processing-in-memory (PIM) reduces data movement by executing near DRAM, but our large-scale characterization on real PIM hardware shows end-to-end performance is often limited by disjoint host–device address spaces that force explicit staging transfers. In contrast, CXL-PIM exposes a unified address space and cache-coherent access at the cost of higher access latency. These opposing interface models create workload-dependent trade-offs that small-scale studies miss. In this work, we present a side-by-side, large-scale comparison of PIM and CXL-PIM using measurements from real PIM and trace-driven CXL modeling, and identify when unified-address access amortizes link latency to overcome transfer bottlenecks, and when tightly-coupled PIM remains preferable. Our results reveal phase and dataset-size regimes that flip the ranking between the two designs and provide practical guidance for near-memory system architects.
\end{abstract}

\maketitle

\begin{figure*}[htb]
        \centering
        \includegraphics[width=\textwidth]{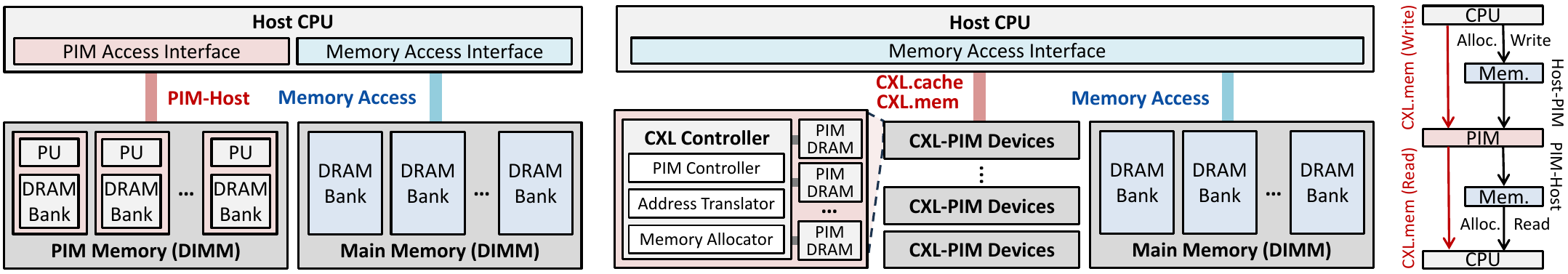}
        \caption{Architecture and dataflow of traditional DIMM-based PIM and CXL-PIM. Traditional PIM requires explicit host–to/from-PIM transfers, while CXL-PIM provides unified memory access through CXL.mem.}
        \label{fig:cxlpim}
\end{figure*}

\section{Introduction}
\noindent
The rapid growth of data-intensive applications such as large language models, graph analytics, and recommendation systems imposes significant pressure on memory bandwidth and capacity. Processing-in-memory (PIM)\footnote{Throughout this paper, we use the term \emph{PIM} to refer specifically to conventional DIMM-based PIM architectures with disjoint host and PIM address spaces.} aims to alleviate this pressure by embedding lightweight processing units within memory modules to reduce host–DRAM traffic~\cite{mutlu2022modern}. This direction has gained substantial industrial traction, with commercial prototypes such as Samsung’s AXDIMM~\cite{ke2021near}, SK Hynix’s AiM~\cite{kwon2022system}, UPMEM’s PIM-DIMMs~\cite{devaux2019true}, and HBM-PIM devices~\cite{kim2021aquabolt, lee2021hardware}. Academic studies further characterize these systems and demonstrate their feasibility on representative workloads~\cite{gomez2021benchmarking}.

In parallel, Compute Express Link (CXL)~\cite{das2024introduction} introduces an alternative near-memory integration model. CXL provides a cache-coherent, byte-addressable interface that exposes device memory in a unified address space to the host. Recent efforts reflect this trend: CENT integrates near-memory compute units into a CXL-attached device to support large-scale inference~\cite{gu2025pim}, while Marvell’s Structera A couples DRAM capacity with embedded compute engines as a commercial CXL-PIM product~\cite{marvell_structera_a2504}. These designs exemplify what we call CXL-PIM, a class of systems that pair PIM-capable memory with CXL to present a unified host–device memory view. This approach stands in contrast to conventional PIM, where host DRAM and PIM memory reside in disjoint address spaces and execution requires explicit staging before and after in-memory computation.

To evaluate these contrasting models under realistic conditions, we first recognize that conventional PIM and CXL-PIM embody two fundamentally different interface philosophies: the former relies on disjoint host–device address spaces with explicit staging, while the latter exposes device memory through a unified, byte-addressable interface. These differences imply distinct data movement paths, coherence behaviors, and performance bottlenecks, raising the question of how each architecture behaves under large-scale workloads. To investigate this, we scale established PIM workloads to much larger datasets and measure performance on commercial PIM hardware. This large-scale characterization reveals behavior that small inputs do not expose. As input and output volumes increase, end-to-end execution becomes dominated by staged transfers between the host and PIM, a direct consequence of disjoint address spaces. Although near-memory compute units offer substantial parallelism, repeated copies into and out of PIM memory degrade performance and stall scalability before all processing units (PUs) are fully utilized.

CXL-PIM occupies a different point in the design space. Its unified address space removes staging copies and allows the host to access device memory directly using standard load/store instructions. This benefit, however, comes with higher per-access latency over the PCIe-based CXL path. The two architectures, therefore, expose distinct bottlenecks rather than a single winner. The relative outcome depends on workload characteristics, data volume, input–output balance, access regularity, and the extent to which transfers can be overlapped with computation. Motivated by this observation, we perform a systematic, large-scale comparison of PIM and CXL-PIM to characterize their architectural trade-offs and to identify when unified-address access yields meaningful benefits and when traditional PIM remains advantageous.

Taken together, these observations necessitate a systematic examination of PIM and CXL-PIM to understand their architectural trade-offs. This work focuses on characterizing how the two architectures behave at large scales. It further identifies when unified-address access effectively addresses the limitations of conventional PIM and when traditional PIM remains advantageous. Our study provides a comprehensive, measurement-driven foundation for future near-memory system design. In summary, this paper makes the following contributions.

\begin{itemize}
    \item We perform the first large-scale characterization of PIM, revealing structural limitations caused by disjoint address spaces that are not visible under existing small-scale evaluations.
    \item We conduct a systematic comparison of PIM and CXL-PIM to evaluate how their different interface models influence end-to-end performance and scalability.
    \item We identify future research opportunities for PIM and CXL-PIM based on empirical insights obtained from large-scale measurements.
\end{itemize}

\section{Background and Analysis} \label{sec:back}
\subsection{Processing-in-memory (PIM) Systems}
\vspace{0.8mm}\noindent\textbf{System Architecture and Dataflow.} 
Processing-in-memory (PIM) integrates lightweight processing units (PUs) within memory modules to reduce costly data movement across the memory hierarchy and mitigate the host–memory bandwidth bottleneck. Among several PIM designs~\cite{ke2021near, kwon2022system, lee2021hardware}, UPMEM is one of the most widely studied platforms~\cite{chen2024updlrm, gomez2021benchmarking, nider2021case}. Figure~\ref{fig:cxlpim} illustrates a typical DIMM-based PIM system in which PIM memory resides alongside conventional main memory. Each PIM module integrates multiple DRAM chips, and each chip embeds several PUs. Every PU includes a private scratchpad (or cache), a local DRAM bank, and an on-module DMA engine that supports intra-module data transfers. This organization enables large-scale parallel execution across many PUs.

PIM systems operate cooperatively with the host CPU, but the execution flow requires several staging steps. As shown on the right side of Figure~\ref{fig:cxlpim}, applications first allocate buffers in host memory and prepare input data in the host address space. The host then explicitly transfers this data into PIM memory through the PIM–host interface before execution begins. After the PUs finish computation, the results are copied back into host memory and written into the allocated region. This explicit movement of inputs and outputs is inherent to DIMM-based PIM because the PIM memory space is not unified with the host address space.

\begin{figure*}[htb]
    \centering
    \begin{subfigure}[t]{\textwidth}
        \centering
        \includegraphics[width=\textwidth]{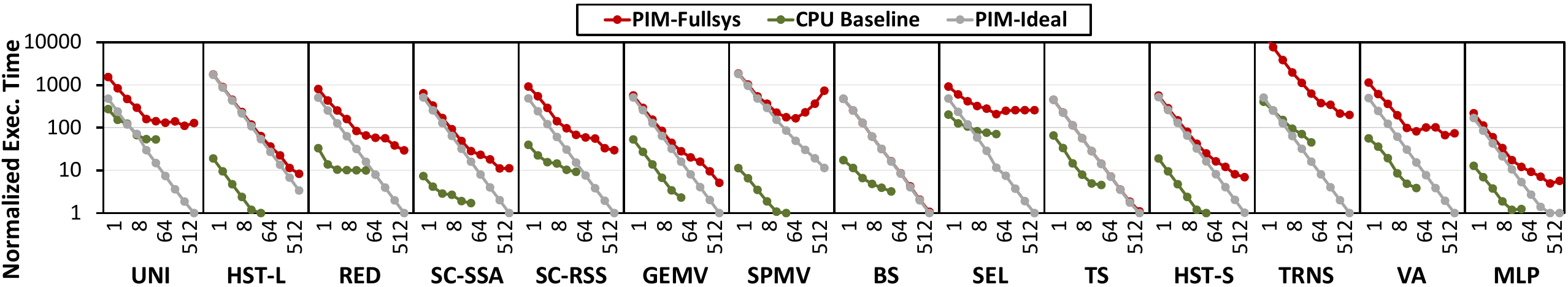}
        \caption{Normalized execution time for CPU baseline, PIM-ideal, and PIM-fullsys. PIM-ideal represents the execution time of PIM processing units only. PIM-fullsys represents the overall end-to-end execution time, including communication overheads between PIM and the host.}
        \label{fig:speedup}
    \end{subfigure}
    \begin{subfigure}[t]{\textwidth}
        \centering
        \includegraphics[width=\textwidth]{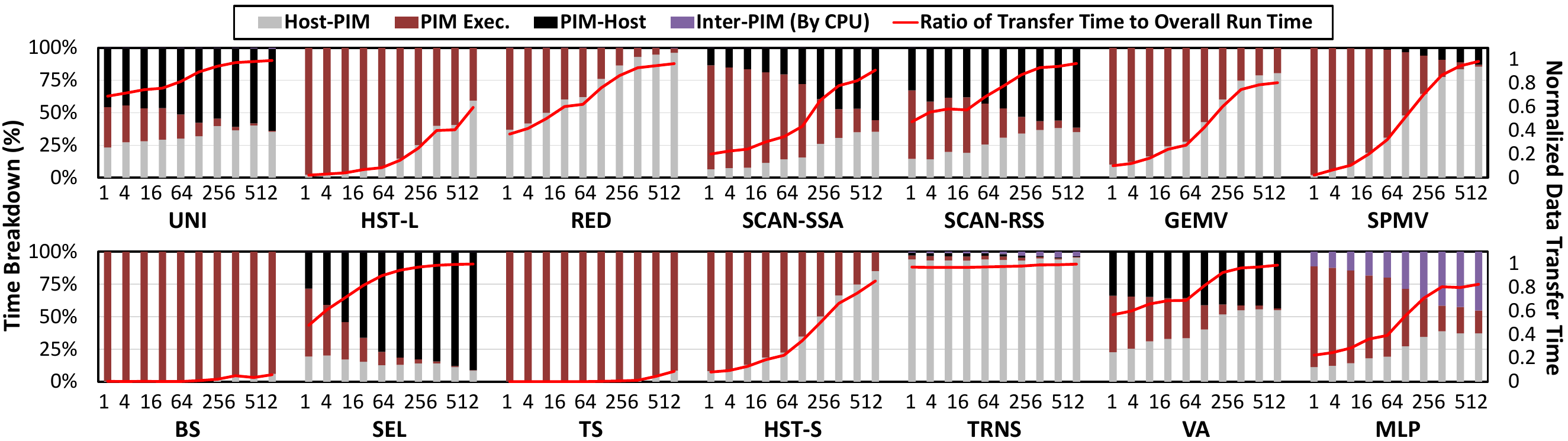}
        \caption{Execution time breakdown of PIM-Fullsys under different workloads. The $X$-axis indicates the number of active PIM processing units (ranging from 1 to 512). The left $Y$-axis shows the percentage of total execution time, while the right $Y$-axis reports the ratio of data transfer time relative to the overall execution time.}
        \label{fig:mov_break}
    \end{subfigure}
    \caption{Overall performance of the PIM system over large-scale workloads, showing that PIM fails to scale as dataset sizes grow.}
    \label{fig:mov}
\end{figure*}

\vspace{0.8mm}\noindent\textbf{Limitations of Current PIM Systems and Benchmarks. } 
Given that DIMM-based PIM requires explicit host–PIM transfers due to its disjoint address spaces, a central question is how severe this overhead becomes at realistic data scales. Existing benchmarks, such as PrIM~\cite{gomez2021benchmarking}, primarily evaluate small datasets and therefore do not capture this behavior. Following the workload definitions in Table~\ref{tab:workloads}, we scale each workload to much larger inputs and evaluate the system under the configuration in Section~\ref{sec:methodology}. Once scaled, the results diverge from prior reports and reveal performance trends that differ fundamentally from those observed in small-scale evaluations.

\begin{table}[ht]
  \centering
  \caption{Workloads~\cite{gomez2021benchmarking} and dataset sizes used in our analysis.}
  \label{tab:workloads}
  \begin{tabular}{|l|c|l|c|}
    \hline
    \textbf{Workload} & \textbf{Dataset Size} & \textbf{Workload} & \textbf{Dataset Size} \\
    \hline
    VA  & 128 GB  & UNI & 128 GB \\
    SEL & 128 GB  & GEMV & 128 GB \\
    SPMV & 634 MB & TS   & 10.24 GB \\
    BS   & 10.24 GB & MLP   & 95.36 GB \\
    SCAN-SSA & 95.36 GB & SCAN-RSS & 95.36 GB \\
    TRNS & 95.36 GB & RED   & 95.36 GB \\
    HST-L & 95.36 GB & HST-S & 95.36 GB \\
    \hline
  \end{tabular}
\end{table}

Using these scaled workloads, we analyze end-to-end PIM behavior under realistic conditions. Figure~\ref{fig:speedup} reports the normalized execution time of three platforms: host execution using only the CPU (\textit{CPU Baseline}), PIM with full-system effects (\textit{PIM-Fullsys}), and PIM excluding data transfer overheads (\textit{PIM-Ideal}). The CPU baseline and PIM are configured with up to 32 cores and 512 PUs across eight ranks, respectively. \textit{PIM-Ideal} reflects the raw compute capability of the PIM architecture. Two observations emerge. First, without transfer overheads, \textit{PIM-Ideal} achieves substantial acceleration: execution time decreases nearly proportionally with the number of PUs and reaches its minimum at 512 PUs. In contrast, when full-system effects (i.e., the staged dataflow in Figure~\ref{fig:cxlpim}) are included, the average end-to-end runtime of \textit{PIM-Fullsys} is often slower than the 32-core CPU baseline. Second, data movement overhead not only slows execution but also scales poorly with increasing PU counts. For workloads such as UNI, SPMV, SEL, and VA, adding more PUs can even increase total runtime.

Figure~\ref{fig:mov_break} further breaks down \textit{PIM-Fullsys} time into four components: host-to-PIM transfers (Host–PIM), PIM execution (PIM Exec.), PIM-to-host transfers (PIM–Host), and inter-PU communication managed by the host (Inter-PIM). Host–PIM data movement dominates, accounting for 60–90\% of the total runtime across most workloads. For example, in VA, SEL, and TRNS, more than 80\% of the time is spent in Host–PIM and PIM–Host transfers, while actual PIM execution contributes less than 15\%. The dominant factor varies by workload: VA is constrained by both input and output transfers due to symmetric data volumes; SEL is limited primarily by PIM–Host transfers because of its large output size; and TRNS is bottlenecked by Host–PIM transfers because a large input must be staged into PIM. Moreover, MLP requires synchronization across all PUs, which leads to heavy inter-PU communication (up to 60\% of total time at 512 PUs) and further amplifies data movement overhead, limiting scalability.

Taken together, these results contrast sharply with conclusions drawn from earlier small-scale benchmarks. When realistic dataset sizes are used, \textit{PIM-Fullsys} not only loses the advantage reported in prior work but can also perform worse than the CPU baseline, with scalability collapsing as PU counts increase. This behavior stems from the fundamental cost of host–PIM transfers rather than from insufficient PIM compute capability. \emph{These limitations directly motivate designs such as CXL-PIM, which seek to eliminate explicit transfers by exposing a unified host–device memory model}.

\subsection{CXL-PIM Systems}
\vspace{0.8mm}\noindent\textbf{Compute Express Link (CXL)} 
Compute Express Link (CXL) is an industry-standard interconnect that offers a cache-coherent, high-bandwidth, and low-latency interface to external devices over PCIe~\cite{das2024introduction}. Among the supported device classes, \emph{CXL Type-2} devices couple accelerator logic with attached memory and map that memory into the host’s address space~\cite{ji2024demystifying, ham2024low, gu2025pim}. In this model, the host CPU can access device memory with standard load/store instructions, avoiding explicit copy operations. Consequently, CXL provides a path to address a key limitation of conventional DIMM-based PIM by eliminating the disjoint host–device address spaces that otherwise mandate costly staging transfers. By attaching a PIM module as a CXL device and integrating it into a unified address space with host DRAM, the CPU can offload computation while directly reading and writing device-resident data. 

\vspace{0.8mm}\noindent\textbf{Existing CXL-PIM Designs} 
Recent systems explore combining near-memory compute with the CXL interface. A representative research system, \emph{CENT}, integrates processing-in/near-memory units into a CXL Type-2 device to support large-scale inference. The host accesses device memory through a unified, coherent interface, and kernels execute close to the memory chips~\cite{gu2025pim}. CENT’s design demonstrates how unified memory access simplifies data sharing between host and near-memory compute and removes explicit staging overheads. On the product side, Marvell’s \emph{Structera A} is a commercial CXL 2.0 near-memory accelerator that couples multi-channel DRAM capacity with embedded compute engines~\cite{marvell_structera_a2504}. These efforts exemplify a practical integration model in which PIM-capable devices attach via CXL and expose a unified host–device memory view.

\vspace{0.8mm}\noindent\textbf{System Architecture and Dataflow}
Figure~\ref{fig:cxlpim} summarizes a generic CXL-PIM organization. Regardless of internal microarchitecture, a CXL-PIM device (1) attaches PIM-capable memory through \texttt{CXL.mem}, (2) exposes device memory within a unified, coherent address space, and (3) allows the host to manage PIM execution without explicit host–device copy APIs. As illustrated on the right of Figure~\ref{fig:cxlpim}, the host allocates and initializes inputs directly in the CXL-PIM memory region. After PIM execution, results are read from the same addresses using ordinary loads. This workflow is more straightforward than the DIMM-based PIM because staging buffers and format conversions are unnecessary. \emph{Despite unified addressing, accesses still traverse a PCIe-based CXL path. Per-access latency is therefore higher than DDR-attached DIMMs. As a result, CXL-PIM removes staging cost but introduces a different bottleneck profile, which we quantify in later sections.}

\section{Motivation}
\noindent
Our large-scale characterization reveals a fundamental limitation of DIMM-based PIM systems: end-to-end performance degrades primarily because of the staged host–PIM transfer model enforced by disjoint address spaces, rather than a lack of compute capability in the PUs.

\vspace{0.8mm}\noindent\textbf{Architectural Trade-offs: PIM or CXL-PIM? } 
PIM and CXL-PIM represent two contrasting integration models for near-memory computation. Conventional PIM benefits from tight coupling to DRAM but requires explicit host–device transfers due to separate address spaces. CXL-PIM removes these staging transfers by exposing a unified address space to the host, although each access traverses a PCIe-based CXL path and therefore incurs higher per-access latency. These differences imply that neither design is universally superior. The favorable choice depends on workload properties such as input volume, output size, access regularity, and the extent to which transfer and computation can be overlapped. Prior studies, however, rarely provide a large-scale side-by-side evaluation that can expose these trade-offs under realistic conditions. This gap motivates our benchmarking-driven analysis to determine how each architecture behaves across regimes and to identify when PIM or CXL-PIM is preferable.

\vspace{0.8mm}\noindent\textbf{Exposing realistic behavior of PIM and CXL-PIM. }
Our objective is to conduct a large-scale evaluation of PIM and CXL-PIM that reveals their architectural behavior under realistic data sizes. At these scales, workloads exercise the complete host–device dataflow, making the effects of input volume, output size, and intermediate transfers observable in end-to-end time. Only under such conditions do the limitations of DIMM-based PIM and the potential benefits of unified access in CXL-PIM become apparent.

The resulting large-scale workloads form the foundation of our study. They allow a controlled comparison of PIM and CXL-PIM under identical configurations. \emph{This thus enables us to determine which kernels benefit from unified addressing and which continue to perform effectively on conventional PIM.} Equally important, they reveal where both architectures fall short, providing insight that motivates architectural refinements such as pipelined execution, device-assisted coordination, and workload-aware memory allocation.

\section{Methodology}
\label{sec:methodology}
\noindent\textbf{PIM hardware configuration.}
Our PIM experiments are conducted on a real PIM system~\cite{devaux2019true}. The host platform consists of two Intel Xeon Silver 4110 CPUs equipped with 128~GB of DDR4 memory. The PIM subsystem comprises four dual-rank PIM-DIMMs, each containing 16 DRAM chips with eight PUs per chip, for a total of 512 PUs. Every PU incorporates a RISC core running at 350\,MHz, supports 24 hardware threads, and includes a 64\,KB private scratchpad and a dedicated 64\,MB DRAM bank. This configuration allows us to observe realistic execution behavior, memory access characteristics, and host–device interactions based on actual hardware operation.

\noindent\textbf{CXL-PIM system implementation.}
We model the CXL-PIM architecture using CXLMemSim~\cite{cxlsimgithub, yangyarch23}, an open-source simulator for characterizing \texttt{CXL.mem} latency and bandwidth behavior. For each CXL-PIM device, we extend the simulator to support rank-level parallelism across DDR channels, enabling all four ranks to independently service requests through multiple channels~\cite{lee2024analysis, ghose2019demystifying}. The modeled CXL-PIM configuration contains up to 256 PUs distributed evenly across four ranks, reflecting the compute organization of a realistic Type-2 CXL device.

To capture representative behavior, workloads are first executed on the real PIM platform, and we record communication between the host and PIM using profiling tools from the UPMEM SDK~\cite{upmem_sdk_profile}. These traces preserve the ordering, granularity, and patterns of host-issued requests. Layout transformations such as data transpose are performed before trace collection on the real PIM hardware, consistent with its required preprocessing behavior. In a CXL-PIM system, however, these transformations can be handled by the CXL controller rather than the CPU and therefore do not incur host-side overhead. To maintain fairness with the real PIM hardware, we incorporate this expected preprocessing cost into the modeled \texttt{CXL.mem} access latency. This yields a deliberately conservative configuration whose effective latency is higher than the commonly used values in prior CXL studies~\cite{ham2024low, li2023pond}.

\begin{table}[t]
\centering
    \caption{System configuration.}
    \begin{tabular}{ll}
        \hline
        \multicolumn{2}{c}{\textbf{Host}} \\
        \hline
        Processor & 2 Intel Xeon Silver 4110 CPUs @ 2.10\,GHz\\
        Standard DIMMs     & 128\,GB DDR4 \\
        \hline
        \multicolumn{2}{c}{\textbf{PIM System~\cite{devaux2019true}}} \\
        \hline
        PIM DIMMs       & 8 ranks; 16 chips/module; 8 PUs/chip \\
        PU & RSIC core @ 350\,MHz; 24 hardware threads  \\
        PU's memory  & 64\,KB scratchpad; 64\,MB DRAM bank \\
        \hline
        \multicolumn{2}{c}{\textbf{CXL-PIM System}} \\
        \hline
        CXL model & CXLMemSim~\cite{yangyarch23, cxlsimgithub} \\
        CXL link & PCIe Gen5 $\times$ 8; 64\,GB/s bidirectional~\cite{das2024introduction}\\
        Switching delay & 70\,ns~\cite{das2024introduction, li2023pond}\\
        CXL.mem latency & 180\,ns (read and write)~\cite{ham2024low, li2023pond} \\
        Device memory tech. & 4 $\times$ DDR4-2400 channels (up to 256 PUs) \\
        \hline
    \end{tabular}\label{tab:system_config}
\end{table}

\label{sec:analysis}
\begin{figure*}[t]
        \centering
        \includegraphics[width=0.99\textwidth]{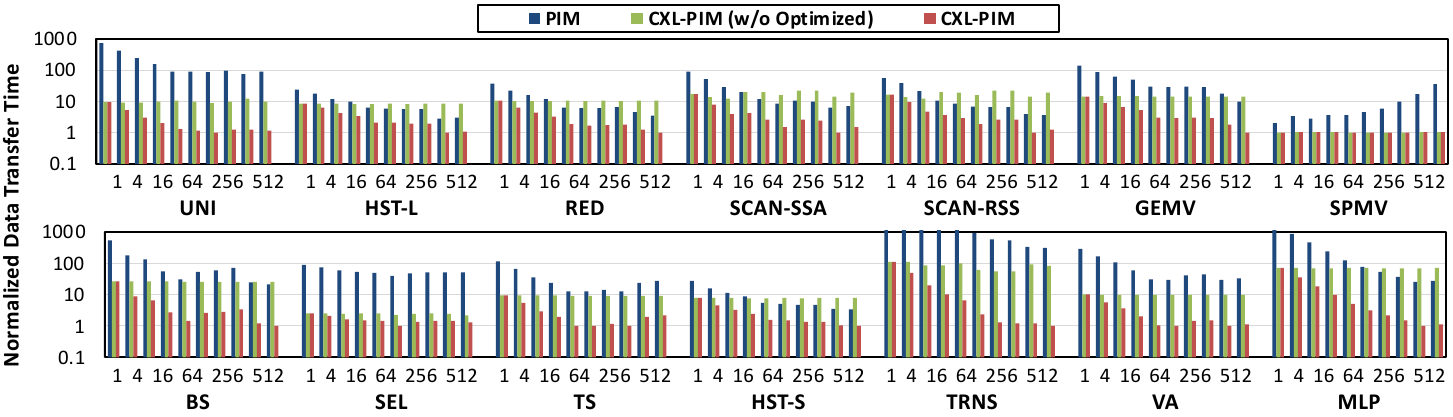}
        \caption{Normalized data transfer time between the host and PIM devices. We fix the overall data size and scale the number of PUs from 1 to 512.}
        \label{fig:exp_tran_time}
\end{figure*}

The simulator enforces link-level constraints based on the modeled CXL interconnect. We configure a PCIe Gen5~$\times$8 link, providing 64\,GB/s bidirectional bandwidth~\cite{das2024introduction}. The \texttt{CXL.mem} access latency is set to 180\,ns for both reads and writes, following empirical values from real CXL-ready systems~\cite{ham2024low, hou2024understanding, sun2023demystifying}. In addition, we include a 70\,ns switching delay for data transfers, consistent with prior system-level characterizations~\cite{das2024introduction, li2023pond}. Table~\ref{tab:system_config} summarizes all configuration parameters for both the PIM and CXL-PIM systems.

\section{Performance Analysis} \label{sec:perf}
\subsection{Experimental Setup}
We scale the input datasets of the PrIM benchmark, which provides representative kernels for commercial PIM devices, to sizes up to 128\,GB. Several PrIM workloads, such as NW and BFS, are excluded due to scalability challenges. For example, enlarging BFS introduces graph partitioning issues that significantly complicate analysis. The workloads and dataset sizes are summarized in Table~\ref{tab:workloads}. Due to implementation constraints, not all workloads can be scaled to the full 128\,GB. For these cases, we use the largest dataset size that executes successfully.

We evaluate three system configurations. The first is the PIM setup (\textit{PIM}) described in Section~\ref{sec:methodology}. The second is the CXL-PIM design, in which PIM is integrated as a CXL device with up to 256 PUs distributed across four ranks. In this configuration, the host can access PIM memory directly through the CXL interface without explicit copy operations (\textit{CXL-PIM}). To isolate the impact of rank-level parallelism inside the CXL-PIM device, we also evaluate a variant without this capability, thereby limiting effective bandwidth utilization across DDR channels (\textit{CXL-PIM w/o Optimized}).

\subsection{Analysis on Data Transfer Overheads}
\noindent
Figure~\ref{fig:exp_tran_time} reports the normalized data transfer time for PIM, CXL-PIM w/o Optimized, and CXL-PIM as we scale the number of PUs. Three key behaviors emerge.

First, CXL-PIM consistently achieves the lowest transfer time across all workloads. At 512 PUs, the improvement over PIM is typically between 9$\times$ and 10$\times$, and exceeds 50$\times$ for UNI, TRNS, SPMV, and MLP. These workloads are dominated by round-trip data movement in PIM, so removing explicit copies through direct memory access on CXL yields disproportionately large benefits. Second, CXL-PIM w/o Optimized improves over PIM only for a subset of workloads, including UNI, SPMV, SEL, TRNS, and VA. The advantage shrinks as the PU count increases. Without parallel handling of requests within the device, \texttt{CXL.mem} traffic effectively serializes on the CXL-PIM side, and transfer time grows with PU-level concurrency, limiting the net gain. Third, comparing CXL-PIM with the unoptimized configuration shows that optimized intra-rank parallel transfers are essential. Most workloads experience an additional 10$\times$ to 20$\times$ reduction in transfer time once multi-request parallelism is enabled. An exception arises in SPMV and SEL, where PU communication patterns cannot be aligned to exploit parallel transfers. In these cases, communication remains effectively sequential per PU, and CXL-PIM cannot scale.

These results clarify the architectural factors that shape CXL-PIM behavior. The effectiveness of CXL-PIM depends strongly on the degree of parallelism inside the CXL-attached DRAM subsystem. The design relies on concurrent transfers across chips and ranks while many PUs issue requests simultaneously. When the PIM-side organization enforces sequential per-PU communication, as observed in HST-L, HST-S, RED, SCAN-SSA, and SCAN-RSS, end-to-end transfer time remains high despite the unified-address interface.

Overall, CXL improves data transfer performance by enabling direct memory access and eliminating explicit round-trip copy operations. These advantages are most pronounced in workloads where PIM computation is short and transfer time dominates overall execution, such as UNI, SPMV, SEL, TRNS, and VA. In contrast, when data movement is not the primary bottleneck, the improvement is smaller because \texttt{CXL.mem} access latency is higher than that of native DDR, and device-side delays can offset the benefit.

\begin{observationbox}
\textit{\textbf{Observation \#1:} Workloads that are tightly constrained by PIM’s transfer model and have very short PU execution time benefit directly from CXL’s unified memory access. When data movement is not the dominant bottleneck, additional system-level optimizations are required to fully expose the performance potential of CXL-PIM.}
\end{observationbox}

\begin{figure*}[t]
        \centering
        \includegraphics[width=0.99\textwidth]{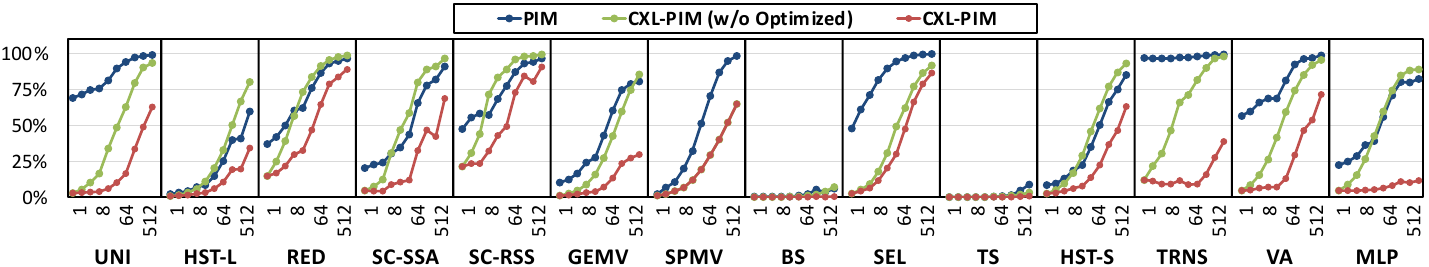}
        \caption{The ratio of data transfer time to the end-to-end execution time. We fix the overall data size, and scale the number of DPUs from 1 to 512.}
        \label{fig:exp_ratio}
\end{figure*}

\subsection{Analysis on Scaling Results and Time Breakdown}
\noindent
As the number of PUs increases, the ratio of data transfer time to end-to-end execution generally grows. The total work is divided across more PUs, so per-PU execution time decreases, whereas transfer cost does not automatically benefit from additional PUs unless there is sufficient communication parallelism. The measured results reveal three distinct groups of workload behaviors.

The first group, TS and BS, naturally maintains a low transfer ratio. Even at 512 PUs, their transfer fraction remains below 0.1 in both PIM and CXL-PIM, indicating that computation dominates execution time across all architectures. These workloads are inherently communication-friendly and do not exhibit scaling-related degradation

The second group, UNI, SPMV, SEL, TRNS, and VA, is heavily constrained by transfer overhead in PIM. For these workloads, PIM consistently shows very high transfer ratios, often exceeding 0.85 at 512 PUs. CXL-PIM substantially reduces this ratio, even without system-level optimization. However, the improvement does not scale linearly with the number of PUs, because the \texttt{CXL.mem} link must still handle growing request concurrency as more PUs issue transfers in parallel.

The third group, HST-L, RED, SCAN-SSA, SCAN-RSS, GEMV, and HST-S, cannot fully exploit CXL’s architectural advantages. Their transfer ratios under PIM also rise sharply, with many workloads exceeding 0.85 at 512 PUs. Under CXL-PIM w/o Optimized, the transfer ratio remains high, typically between 0.80 and 0.99. This is because communication patterns are effectively sequential per PU and cannot be parallelized inside the CXL device. Even under optimized CXL-PIM, the transfer ratio decreases only to around 0.5 for many of these workloads, indicating that they require more aggressive device-side parallel memory access or restructured communication patterns to benefit significantly.

\begin{observationbox}
\textit{\textbf{Observation \#2:} As PU count increases, CXL-PIM w/o Optimized exhibits the steepest growth in transfer ratio, whereas optimized CXL-PIM shows a similar but more moderate trend. Workloads such as MLP, which involve intensive bidirectional communication, benefit significantly once device-side parallelism is enabled, demonstrating markedly improved scaling behavior under optimized CXL-PIM.}
\end{observationbox}

\section{Research Opportunities and Directions}
\label{sec:future}
\noindent
Our analysis in Section~\ref{sec:perf} shows that CXL-PIM does not inherently benefit every workload, nor does it always scale as effectively as DIMM-based PIM. Although CXL-PIM removes the disjoint-address–space bottleneck, its scalability is still constrained by device-side coordination overhead and limitations in data movement efficiency. These findings indicate that additional architectural and system-level optimizations are required for CXL-PIM to reach its full potential.

\subsection{Pipelining PU Execution and Transfer}
\noindent
Existing DIMM-based PIM systems are unable to effectively overlap data transfers with computation~\cite{lee2024analysis}. By contrast, \emph{CXL introduces an opportunity to pipeline execution and communication.} Instead of serializing data movement and PU computation, future designs can offload upcoming data batches to the CXL switch or PIM device while PUs are still processing. The CXL-PIM controller can stage data during execution and deliver it as soon as PUs are ready; likewise, when PUs begin processing the next batch, results from the previous batch can be transferred back to the host in parallel. Such pipelining reduces idle cycles and improves efficiency, particularly for large-scale workloads.

\begin{figure}[t]
        \centering
        \includegraphics[width=0.49\textwidth]{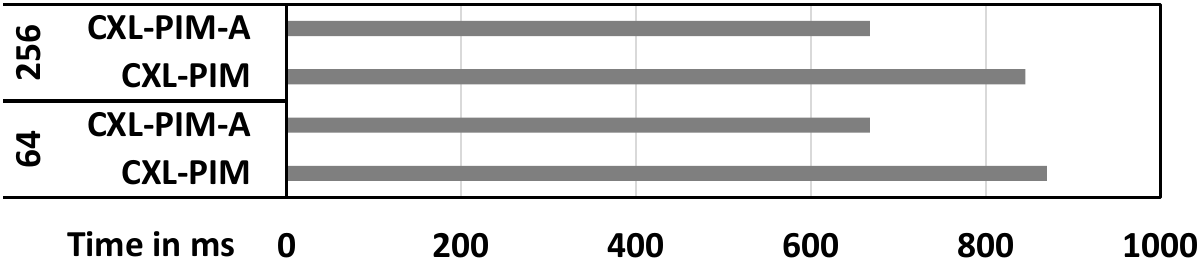}
        \caption{Data transfer times of CXL-PIM w/ and w/o CXL-Assisted PU Management under MLP workload.}
        \label{fig:exp_cxla}
\end{figure}

\subsection{CXL-Assisted Data Flow Management}
In current systems, the CPU is responsible for PU scheduling and coordination, introducing significant overheads and additional data transfers. \emph{With CXL, much of this responsibility can be delegated to the device-side controller, or even to the CXL switch.} Allowing the controller to manage PU execution and data delivery reduces host involvement and minimizes unnecessary round-trips. For example, in workloads such as MLP or other ML tasks~\cite{gomez2022machine} that require multiple computation rounds with intermediate aggregation, controller-assisted management combined with workload-aware data placement can substantially reduce inter-PU communication overhead.
As shown in \figurename~\ref{fig:exp_cxla}, enabling CXL-Assisted Data Flow Management (CXL-PIM-A) achieves more than $1.3\times$ performance improvement over baseline CXL-PIM, since the host CPU is no longer involved in inter-PU coordination and all communication remains within the device. Importantly, this result reflects only the direct reduction in communication overhead. With additional support from the CXL controller, such as combining pipelined execution with coordinated transfers, even greater efficiency can be achieved, underscoring the strong potential of device-assisted management in future CXL-PIM systems.


\subsection{Memory Allocation Policy}
\noindent
Data allocation across CXL-PIM devices is another critical factor for efficiency. Cooperating PUs and their associated data should be placed within the same device to minimize communication latency. For example, MLP workloads that require frequent inter-PU communication benefit from keeping all related data within a single device. By contrast, workloads with independent computation units can distribute PUs and data across multiple devices to exploit aggregate bandwidth. For instance, VA workloads can be partitioned into independent units. Developing workload-aware allocation strategies that balance coordination with bandwidth utilization is, therefore, essential to fully harness the potential of CXL-PIM.

\section{Related Works}
\noindent\textbf{Communication Issues in PIM Systems.}
Several recent efforts attempt to reduce the host–PIM transfer overhead. PIM-MMU~\cite{lee2024pim} introduces a hardware–software co-design that offloads data movement to a dedicated copy engine and employs PIM-aware scheduling, which improves throughput but remains constrained by the disjoint address space model. Similarly, Pid-comm~\cite{noh2024pid} provides lightweight communication primitives to better coordinate data movement among PIM PUs, though it also relies on explicit transfers across separate memory spaces. PIMnet.~\cite{son2025pimnet} proposes an interconnection network to enable efficient collective communication among PIM banks, thereby addressing the limitations of inter-PU communication.

\vspace{0.6mm}\noindent\textbf{CXL–PIM Systems.}
Recent work has also explored integrating PIM with CXL. CENT~\cite{gu2025pim} presents a CXL-enabled system for LLM inference that leverages GDDR6-based PIM to eliminate the need for GPUs. Toleo~\cite{dong2024toleo} demonstrates a secure CXL-based smart memory design that scales freshness guarantees to terabyte-scale CXL-expanded memory. CXL-PNM~\cite{park2024lpddr} develops a processing-near-memory platform based on LPDDR that addresses the inefficiencies of competing technologies such as HBM-PIM and AxDIMM.
Ham et al.~\cite{ham2024low} propose a general-purpose near-data processing architecture for CXL memory expanders, but the evaluation remains simulation-based rather than using real PIM traces.
Intelligent Knowledge Store~\cite{quinn2025compute} introduces a CXL memory expander with near-memory accelerators for retrieval-augmented generation.

\section{Conclusion}
\noindent 
This paper revisits the fundamental data movement bottleneck in commercial PIM systems. Through large-scale characterization, we show that while PIM cores provide massive raw compute capability, end-to-end performance is dominated by host–PIM transfers. By introducing and evaluating CXL-PIM, we expose PIM as CXL devices, thereby providing a unified memory address space, removing redundant data copies, and enabling scalable host–PIM cooperation. Our results demonstrate that optimized CXL-PIM designs achieve significant scalability improvements for large-scale data-intensive workloads.

\bibliographystyle{ACM-Reference-Format}
\bibliography{reference}

\end{document}